\documentclass[12pt,preprint]{aastex}

\shorttitle{Evidence for a companion to BM~Gem}
\shortauthors{Izumiura, Noguchi, Aoki et al.}

\begin{document}

\title{Evidence for a companion to BM~Gem, a silicate carbon star\footnote{
Based on data collected at the Subaru Telescope, which is operated by
the National Astronomical Observatory of Japan}}


\author{Hideyuki Izumiura\altaffilmark{2,10},
Kunio Noguchi\altaffilmark{3},
Wako Aoki\altaffilmark{3},
Satoshi Honda\altaffilmark{3},
Hiroyasu Ando\altaffilmark{3},
Masahide Takada-Hidai\altaffilmark{4},
Eiji Kambe\altaffilmark{5},
Satoshi Kawanomoto\altaffilmark{6},
Kozo Sadakane\altaffilmark{7},
Bun'ei Sato\altaffilmark{8,2},
Akito Tajitsu\altaffilmark{9},
Wataru Tanaka\altaffilmark{3},
Ki'ichi Okita\altaffilmark{2},
Etsuji Watanabe\altaffilmark{2},
and Michitoshi Yoshida\altaffilmark{2}
}


\altaffiltext{2}{Okayama Astrophysical Observatory, National Astronomical Observatory,
 Kamogata, Asakuchi, Okayama 719-0232, Japan}
\altaffiltext{3}{Division of Optical and Infrared Astronomy, National Astronomical Observatory,
 Mitaka, Tokyo 181-8588, Japan}
\altaffiltext{4}{Liberal Arts Education Center, Tokai University, Hiratsuka,
 Kanagawa 259-1292, Japan}
\altaffiltext{5}{Department of Earth and Ocean Sciences, National Defense Academy,
 Yokosuka, Kanagawa 239-8686, Japan}
\altaffiltext{6}{Astronomical Data Analysis Center,
 National Astronomical Observatory, Mitaka, Tokyo 181-8588, Japan}
\altaffiltext{7}{Astronomical Institute, Osaka Kyoiku University, Kashiwara,
 Osaka 582-8582, Japan}
\altaffiltext{8}{Graduate School of Science and Technology, Kobe University,
 Kobe, Hyogo 657-8501, Japan}
\altaffiltext{9}{Subaru Telescope, National Astronomical Observatory of Japan,
 650 North A'ohoku Place, Hilo, HI 96720, USA}
\altaffiltext{10}{
e-mail: izumiura@oao.nao.ac.jp
}

\begin{abstract}
Balmer and Paschen continuum emission as well as Balmer series lines of
P Cygni-type profile from H$_{\gamma}$ through H$_{23}$ are revealed in
the violet spectra of BM~Gem, a carbon star associated with an oxygen-rich
circumstellar shell (``silicate carbon star'') observed with the high 
dispersion spectrograph (HDS) on the Subaru telescope.
The blue-shifted absorption in the Balmer lines indicates the presence of
an outflow, the line of sight velocity of which is at least 400 km~s$^{-1}$,
which is the highest outflow velocity observed to date in a carbon star.
The Balmer lines showed a significant change in profile over a period of 75 days.
Strong \ion{Ca}{2} K emission was also detected, while \ion{Ca}{2} H emission,
where H$\varepsilon$ overlapped, was absent on both observation occasions.
Violet spectra of the other two silicate carbon stars, V778~Cyg and EU~And, 
and of the prototypical J-type carbon star, Y~CVn, were also observed,
but none of these were detected in either continuum emission below 4000 {\AA}
or Balmer lines.
We argue that the observed unusual features in BM~Gem are strong evidence
for the presence of a companion, which should form an accretion disk that
gives rise to both an ionized gas region and a high velocity, variable outflow.
The estimated luminosity of $\sim$0.2 (0.03-0.6) $L_{\odot}$
for the ionized gas can be maintained by a mass accretion rate to a dwarf
companion of $\sim10^{-8}~M_{\odot}$~yr$^{-1}$,
while $\sim10^{-10}~M_{\odot}$~yr$^{-1}$
is sufficient for accretion to a white dwarf companion. These accretion rates are
feasible for some detached binary configurations on the basis of
the Bond-Hoyle type accretion process.
Therefore, we concluded that the carbon star BM~Gem is in a detached binary
system with a companion of low mass and low luminosity.
However, we are unable to determine whether this companion object is a dwarf or 
a white dwarf, although the gas outflow velocity of 400~km~s$^{-1}$ as
well as the non-detection in the X-ray survey favor its identity as a dwarf star.
The upper limits for binary separation are 210~AU and 930~AU for
a dwarf and a white dwarf, respectively, in the case of circular orbit.
We also note that the observed features of BM~Gem
mimic those of Mira ($o$~Cet), which may suggest actual similarities in their
binary configurations and circumstellar structures.
\end{abstract}


\section{INTRODUCTION}
When low- and intermediate-mass stars evolve along the asymptotic giant branch
where double shell burning of He and H takes place in the interior, the He burning
becomes unstable and gives rise to periodic thermonuclear runaway (``thermal
pulse'' or ``He shell flash''), which induces mixing of newly synthesized $^{12}$C
and other processed materials at the surface (third dredge-up, Iben 1975; Sugimoto \& Nomoto 1975).
The mixing gradually enhances the surface abundance of carbon, which will
eventually turn a star originally oxygen-rich in the surface chemical
composition into a carbon star (cf. Iben \& Renzini 1983).

Among cool luminous carbon stars, there are a group of stars that
show silicate dust emission features in the mid-infrared at 10 and 18~$\mu$m
(Little-Marenin 1986; Willems \& de Jong 1986); these are the so-called 
``silicate carbon'' stars. Silicates are the signature
of oxygen-rich chemistry in their circumstellar dust shells, while their
optical spectra dominated by molecular absorption bands of C$_{2}$ and CN
show that their atmospheres are carbon-rich.
The oxygen-rich chemistry in their circumstellar envelopes is also
confirmed in the gas-phase by the detection of water vapor masers at 22~GHz
(Benson \& Little-Marenin 1987; Nakada et al. 1987).
Even carbon-rich objects showing crystalline silicate features have been
discovered (Waters et al. 1998; Molster et al. 2001).

The intriguing co-existence of oxygen-rich and carbon-rich chemistries in a single 
system prompted two hypothetical explanations. One was the
binary system consisting of a carbon star and a dust-enshrouded M-type giant 
(OH/IR) star (Benson \& Little-Marenin 1987; Little-Marenin, Benson,
\& Dickinson 1988). The other was more interesting, i.e., that we are witnessing
a brief evolutionary stage where the star is in transition from an oxygen-rich
star to a carbon star by the third dredge-up while the remnant oxygen-rich dust
shell is still visible (Willems \& de Jong 1986, 1988).
These earlier pictures were, however, discarded based on the
absence of both spectroscopic signatures of a luminous M-type (OH/IR)
companion (Noguchi et al. 1990; Lambert, Hinkle \& Smith 1990)
and variability
of the color-indices and silicate features (Chan \& Kwok 1988; Lloyd-Evans 1990)
as well as a stringent requirement that primary and secondary must have
very similar masses (Lambert et al. 1990).
Near-infrared speckle interferometry as well as water maser monitoring also
ruled out the presence of a luminous M-type companion to the primary
carbon star (Engels \& Leinert 1994) among silicate carbon stars.

Morris (1987) suggested that a binary system of a mass-losing red giant
star and either a main-sequence dwarf or a white dwarf companion  can
develop an accretion disk around the companion, sometimes a circumbinary disk,
and even a circum-primary disk, depending on the system configuration.
Lloyd-Evans (1990) was inspired by Morris's picture to
propose that in silicate carbon stars the oxygen-rich material 
accumulated in a disk around a hypothetical companion when the
primary was an oxygen-rich mass-losing star, which later turned
into a carbon star through the third dredge-up.
This picture reconciles the observed red infrared color and
the relatively small extinction in the optical. The disk should
be somewhat thickened and may extend to the circumbinary region
to provide a sufficient mid-infrared flux.
Engels \& Leinert (1994) inferred for V778~Cyg and EU~And
a minimum radius of the molecular reservoir where water masers reside
of 45~sin~$i$~AU, where $i$ denotes the inclination of the reservoir,
which is 90$^{\circ}$ when seen edge-on, based on the assumed
mass of 1~$M_{\odot}$ for the primary,  Keplerian motion for the
maser components, and the constancy of radial velocities of the maser lines.
Kahane et al. (1998) and Jura \& Kahane (1999)
proposed the existence of a circumbinary reservoir in Keplerian motion
in silicate carbon stars, BM~Gem and EU~And, on the basis of
the detection of CO J=1-0 and J=2-1 emission with very narrow widths
at their systemic velocities. They suggested that the silicate
grains reside in the reservoir, which was presumably built by the
influence of a postulated unseen companion when the primary was
an oxygen-rich giant.
Waters et al. (1998) noted the similarity to silicate carbon
stars of Red Rectangle that possesses a circumbinary disk showing
crystalline silicate features, an extended carbon-rich outflow,
and narrow CO emission lines.
However, Yamamura et al. (2000) suggested a picture similar
to that reported by Lloyd-Evans (1990) based on analysis of the ISO spectrum
of another silicate carbon star, V778~Cyg.
They found that dust grains in the circumbinary region where the
grains can be warm enough to emit the silicate features will be blown
out in less than one orbital period, and thus unable to form a circumbinary
reservoir. They concluded that the source of the silicate features must be
the oxygen-rich material continuously blown out from the disk around
a companion by the primary's wind and radiation pressure.

The above scenarios all postulate a low-mass, low-luminosity companion,
although no observational evidence has yet been provided.
To gain further understanding of silicate carbon stars, it is essential to
determine whether they indeed have a companion star. It would be very difficult,
however, to detect a postulated low-luminosity companion in the optical or
longer wavelength bands because the primary carbon star must outshine
the companion by many orders of magnitude in these wavelength regions.
In this respect, it has long been known that cool carbon stars 
exhibit extreme violet flux deficiency (Shane 1928), the agent of which
has not yet been unambiguously identified (``violet opacity problem'').
Indeed, cool carbon stars are very dim in the violet region;
e.g., $U-V$ of the carbon star Y~CVn is 8.9 (Nicolet 1978), while those of 
M5 giants and dwarfs are 4.2 and 2.8, respectively (Allen 1976).
This phenomenon can be exploited to search for signatures of companions
to silicate carbon stars in the violet spectral region.
Therefore, we performed high sensitivity spectroscopic observations of 
the visually brightest silicate-carbon stars BM~Gem, V778~Cyg, 
and EU~And, in the violet spectral region.
While we saw no significant violet emission in V778~Cyg or EU~And,
we have detected a featureless continuum below 4000 {\AA} in BM~Gem,
where the Balmer continuum is higher than the Paschen continuum at the Balmer limit,
which is very unusual for a cool luminous carbon star.
In addition, the Balmer series lines were traced from H$_{\gamma}$ up to H$_{23}$
and they showed distinct P Cygni-type profiles. The profiles give a gas
expansion velocity of at least 400 km~s$^{-1}$. Such a high velocity outflow
has never been observed in any type of currently mass-losing carbon star.
Additional observations 75 days later further revealed considerable time
variability of the Balmer lines.
In addition, the BM~Gem system was found to be similar to the Mira ($o$~Cet) system,
which is the prototype of a binary system consisting of a luminous
AGB star (Mira~A) and a low-mass, low-luminosity companion (Mira~B).
In the following sections, we discuss the origin of the continuum emission
and the P Cygni-type Balmer lines, and argue for the presence of a companion to BM~Gem.

\section{OBSERVATIONS}
Observations of BM~Gem, the brightest silicate carbon star in the 
visual region, were made with the high dispersion spectrograph 
(HDS; Noguchi et al. 2002) at the Nasmyth focus of the Subaru telescope 
(Kaifu et al. 2000; Iye et al. 2004) on January 29 and April 14, 2001.
We used the atmospheric dispersion corrector for the optical Nasmyth focus.
The entrance slit width was set to 360~$\mu$m ($0\farcs72$), which corresponded
to a resolution of $\sim$50,000. A quartz glass filter was inserted behind
the slit. HDS is equipped with two mosaiced CCDs from
EEV with 2048$\times$4100 pixels of 13.5~$\mu$m square.
We used the 250~grooves~mm$^{-1}$ cross-disperser grating to observe
the range between 3550--5200 {\AA}  (3550--4350 {\AA} and 4400--5200 {\AA}).
We made two 30-minute exposures and one 30-minute
exposure in January and April, respectively.
A bright J-type carbon star, Y~CVn,
was also observed for 15 minutes on February 1, 2001 (UT) with the same settings
for comparison, because all the silicate carbon stars examined
spectroscopically were known to be J-type stars (Lloyd-Evans 1990),
which have low $^{12}$C/$^{13}$C abundance ratios ($\lesssim 10$) in the
atmosphere (Bouigue 1954).
We also observed the second and third brightest silicate carbon stars,
V778~Cyg and EU~And, for 40 minutes each on July 30, 2001 with almost
the same settings but we employed a slit width of $1\farcs0$ and
2$\times$2 binning for the CCD readout to attain as high a signal to
noise ratio as possible, which gave an effective spectral resolution of
$\sim$38,000. A summary of the observations is given in Table 1.
The last column shows the signal to noise ratio at around 4000 {\AA} in
the reduced one-dimensional spectrum.
The photometric and astrometric data of the targets taken from the
literature are summarized in Table 2. The second and third columns show
the Hipparcos parallax and the uncertainty in the parallax (ESA 1997), respectively.
Near infrared data in the sixth through tenth columns are from Noguchi et al. (1981)
for BM~Gem and Y~CVn and from Noguchi et al. (1990) for V778~Cyg and EU~And.
$B$ and $V$ magnitudes are from the Hipparcos and Tycho catalogs (ESA 1997) for
BM~Gem and Y~CVn, and from Alksnis \& \u Zaime(1993) for V778~Cyg and EU~And.
Data reduction was performed using the echelle package on NOAO IRAF\footnote{
IRAF is distributed by the National Optical Astronomy Observatories, which is
operated by the Association of Universities for Research in Astronomy, Inc. under
cooperative agreement with the National Science Foundation}.
Standard procedures were followed:  bias subtraction, flat fielding, scattered
light subtraction, extraction of one-dimensional spectra,
dispersion correction by Th-Ar lamp spectra,
and removal of the echelle blaze profile by continuum lamp spectra.
Strong cosmic ray events were removed manually.

We detected significant emission in BM~Gem throughout the observed
wavelength range. We then attempted to apply flux calibration
to the observed spectra of BM~Gem.
A star, Feige~34, observed on January 28, 2001 under $\sim0\farcs6$ seeing
with $4\farcs0$ slit width to let virtually all the incident light enter the spectrograph,
was used for our approximate flux calibration.
We derived the system response from the observed spectra
of the hot white dwarf Feige 34. We applied an atmospheric extinction
correction based on the Mauna Kea extinction curve
(Beland, Boulade \& Davidge 1988) to the spectra, and compared them with
the calibrated magnitudes at 50~{\AA} interval of Feige~34 prepared in IRAF,
which is based on Massey et al. (1988).
We then corrected the spectra of BM~Gem for the atmospheric extinction
and the system response.
Finally, we applied a correction for light losses from the entrance slit due to seeing.
Seeing size was measured to be $0\farcs80$' and $0\farcs85$ for January and
April observations, respectively.
We approximated the seeing image by a single Gaussian and obtained
values of 0.71 and 0.68 for the slit transmission efficiency for January and
April, respectively, against $0\farcs72$ slit width assuming perfect telescope guiding.
The spectra of Y~CVn were also calibrated in the same manner for comparison.
The seeing size at the observation of Y~CVn was $0\farcs80$, which gave
a slit efficiency of 0.71.

The uncertainty in the absolute flux densities thus obtained is non-negligible because
the data were taken on different nights with somewhat different seeings
with a slit width close to the size of the seeing image, a flux standard was observed
on only one night, and the violet region is near the atmospheric cutoff, which
is sensitive to extinction corrections.
The uncertainty due to seeing and guiding, which applies to BM~Gem, is estimated
to be $\pm$0.25 mag in each spectrum, adopting maximum possible errors of
$0\farcs2$ in both seeing estimate and guiding. 
The uncertainty in the extinction correction which applies to both BM~Gem and
Feige~34 can give rise to an underestimate in their brightness of about 0.2 mag.
It is obtained by assuming the employed extinction coefficient of
0.37 mag~airmass$^{-1}$ at 3600{\AA} (Beland et al. 1988), with smaller values
for longer wavelengths, was subject to a possible increase of 50\%, although all
our observations were made under clear sky conditions and the airmasses were only
1.01, 1.20, and 1.10 for BM~Gem(Jan), BM~Gem(Apr), and Feige~34, respectively.
Taking all these uncertainties into account, each flux density derived
is likely to be accompanied by an uncertainty range of (-0.45, +0.25),
while the relative uncertainty between BM~Gem(Apr) and BM~Gem(Jan)
can be as large as 0.7(=0.25+0.25+0.2). 
Any difference between the two observations smaller than this magnitude is
dominated by a calibration error and should be treated as insignificant.

We did not apply the above flux calibration to either V778~Cyg or EU~And
because we did not see any significant emission in the region
shortward of 4000 {\AA}. These stars  are 3 magnitudes fainter than BM~Gem
in $B$ (Table 2). If we put BM~Gem farther away from us so that
its $B$ becomes 3 mag fainter, then any signal of the violet
continuum on the raw CCD image becomes only one-quarter or less of
the CCD read noise and merely equal to it or less even in the case of
2$\times$2 binning.
Then, the non-detection of the violet continua here in V778~Cyg and EU~And
indicates that they are not as bright as that of BM~Gem
for their $B$ magnitudes. Nevertheless, it does not necessarily mean
that they do not have a similar violet continuum.
it is necessary to achieve much higher sensitivities for V778~Cyg and EU~And to
distinguish between the presence and absence of a continuum like that
in BM~Gem.
We will not discuss the violet spectra of V778~Cyg or
EU~And further in this paper.

The observed spectrum was corrected for the Doppler shift
and transformed to the wavelength scale with respect to the
local standard of rest (``LSR")
using the tasks RVCORRECT and DOPCOR in IRAF.
The stellar systemic velocity with respect to LSR was also subtracted
in the final spectra shown in this paper. Thus, matter stationary
to the center of mass of the system should appear at zero velocity.
We employed the systemic velocity of 73.2 km~s$^{-1}$ (V$_{\rm LSR}$) for
BM~Gem based on mm-wave CO emission lines
(Kahane et al. 1998; Jura \& Kahane 1999).
This is in agreement with V$_{\rm LSR}$ of 74.7~km~s$^{-1}$
derived from the heliocentric radial velocity of 85.3~$\pm$0.4 km~s$^{-1}$
obtained for the CO first overtone bands by Lambert et al. (1990).
For Y~CVn, we used 21.2~km~s$^{-1}$, based on the mm-wave CO J=1-0 emission
(Izumiura, Ukita, \& Tsuji 1995), as the systemic velocity with respect to LSR.
It should be noted that radial velocities of observed spectral lines given below
will carry a typical uncertainty of $\sim$0.5~km~s$^{-1}$ on the basis of
the observed line widths and strengths, achieved S/N ratio, and employed spectral
resolution.

\section{RESULTS}
\subsection{Violet Spectra of BM~Gem}
The resulting spectra of BM~Gem taken with the CCD that observes the
shorter wavelength portion, reduced to a resolution of 1 {\AA}, are shown
in Figure 1 along with those of Y~CVn.
We have adjusted the vertical scales so that the slopes of the spectra between
4050 {\AA} and 4200 {\AA} look similar to one another.
The spectra of BM~Gem show significant Balmer and Paschen continuum
emission in the violet region, which is evident when compared with Y~CVn,
which shows no detectable violet continuum. What is more intriguing
is that the level of the Balmer continuum is 1.3 times higher than that
of the Paschen continuum when compared around the Balmer limit
in both observation occasions.
This indicates that the continuum emission comes from an ionized gas region.
They also show a series of emission lines that are identified
with Hydrogen Balmer series lines from H$_{\gamma}$ through H$_{23}$.
Moreover, the Balmer lines show P Cygni-type line
profiles, suggesting the presence of an outflow being accelerated
outward against the central continuum source.
H$_{\beta}$ emission was not identified at all in our spectra
recorded on the CCD that covers the longer wavelength part,
probably because the spectral range is dominated by the strong
carbon star spectra.
Broad \ion{Ca}{2} K emission with a blueward depression was also detected
but \ion{Ca}{2} H emission was absent. The \ion{Ca}{2} H emission could be absorbed
by the hydrogen in the outflow because the expected wavelength range
of \ion{Ca}{2} H line (3968.470 {\AA}, Moore 1959) coincides with the absorption core
of the P Cygni profile of H$\varepsilon$ line (3970.074 {\AA}, Moore 1959).
Note here that the features that mimic emission lines, marked with small
vertical ticks seen shortward of 4150 {\AA} in Y~CVn, are contaminations
of ghost spectra of very strong red light of the star, probably due to the 
cross-disperser grating.
The same ghost features are also present in the spectra of BM~Gem at
the same positions, by which only H$_{10}$
of the Balmer lines is significantly contaminated.

The fluxes per unit wavelength at wavelength $\lambda$, $F_{\lambda}$,
of the Balmer continuum of BM~Gem after the approximate flux calibration in \S 2
are 3.0$\times$10$^{-15}$ erg~cm$^{-2}$~s$^{-1}$~{\AA}$^{-1}$
in the January spectra and 4.3$\times$10$^{-15}$ erg~cm$^{-2}$~s$^{-1}$~{\AA}$^{-1}$
in the April spectra immediately shortward of the Balmer limit at 3646 {\AA}.
These correspond to 16.1 mag and 15.7 mag following the definition of
$m\equiv -2.5$~log($F_{\nu}$) $-$ 48.59, where $m$ is the magnitude,
$\nu$ is frequency, and $F_{\nu}$ is the flux per unit frequency at $\nu$
in erg~cm$^{-2}$~s$^{-1}$~Hz$^{-1}$.
Here, $F_{\nu}$ is calculated by $F_{\nu} = F_{\lambda}~|d{\lambda}/d{\nu}|
= c\nu^{-2}F_{\lambda}$ (i.e., $\nu F_{\nu} = \lambda F_{\lambda}$),
where $c$ is the speed of light.
This definition gives 0.048 for the magnitude of Vega at 5556 {\AA} (Hayes \& Latham 1975).
The apparent difference in the continuum level of 0.4 mag is not significant,
considering the possible uncertainty of 0.7 mag given in \S 2.
Thus, we argue that the continuum level is unchanged, and adopt a geometrical mean
of the two values,
3.6$\times$10$^{-15}$ erg~cm$^{-2}$~s$^{-1}$~{\AA}$^{-1}$ or 15.9 mag.
We take the differences of $\pm$0.2 mag between this 15.9 mag
and the two direct values as an additional uncertainty. Adding this to that
given in \S 2, the total uncertainty range has a magnitude of (-0.65, +0.45).
Similarly, the Paschen continuum at 4000 {\AA} has 16.2 mag and 15.6 mag in
the January and April spectra, respectively, and they give 15.9 and (-0.75, +0.55)
in magnitude as the flux level and uncertainty range, respectively.
Note, however, that BM~Gem shows light variation of 0.2--0.4 mag in 100 days in
$B$ and $V$ bands (Alksnis \& \u Zaime 1993). Whether BM~Gem is indeed
variable in the violet continuum is an interesting issue for future studies.

In the original high-resolution spectra of BM~Gem, CN red system lines between
4830--5250 {\AA} give average radial velocities around $-1$~km~s$^{-1}$
with respect to the systemic velocity adopted in \S 2
in the two observations.
Absorption features of \ion{Mn}{1} triplet at 4030-4034 {\AA} were identified,
while low-excitation lines of \ion{Ti}{1} were observed in emission shortward of 4040 {\AA},
both of which are typical for late-N type carbon stars, as described by Gilra (1976).
The \ion{Mn}{1} lines at the two epochs
give average radial velocities around $-6$~km~s$^{-1}$, 
which are compatible with the mm wave CO outflow velocity of
$7.5$~km~s$^{-1}$ and may suggest that they form in
the mass-loss flow of the carbon star. The \ion{Ti}{1} lines have
apparent FWHMs of $\sim$10~km~s$^{-1}$
and show average radial velocities near 0~km~s$^{-1}$ in the two observations,
suggesting that they originate in the extended atmosphere of the carbon star.
The uncertainties in the average radial velocities given above are
as large as 1~km~s$^{-1}$ for all of the CN red system, \ion{Mn}{1} triplet,
and \ion{Ti}{1} lines.  Whether this magnitude of uncertainties,
which is slightly larger than the internal uncertainties of
$\sim0.5$~km~s$^{-1}$ given in \S 2, is due to their intrinsic variabilities
or not is an issue for a future work. 
The continuum shortward of 4040 {\AA} is almost featureless except for
the Balmer and \ion{Ti}{1} lines. No absorption features typical of stellar
photospheres are seen.

Figure 2 shows higher resolution views of the vicinity of H$_{\delta}$,
H$_{\varepsilon}$, and H$_{8}$ lines together with \ion{Ca}{2} K line observed
on the two nights 75 days apart.
The zero systemic velocity corresponds to the middle of the P Cygni profile,
which argues for the outflowing gas being associated with
BM~Gem. The line profiles show that the emitting gas, which is likely
ionized, is expanding at a velocity as large as 400 km~s$^{-1}$
in the sight line. This is greater than the highest velocity outflows observed to date 
in AGB and post-AGB stars, V Hya (200 km~s$^{-1}$, Knapp, Jorissen, \& Young 1997),
CRL 618 (200 km~s$^{-1}$,  Cernicharo et al. 1989),
OH231.8+4.2 (330 km~s$^{-1}$, Alcolea, Bujarrabal, \& S\'anchez Contreras 1996),
and R Coronae Borealis stars (200--350 km~s$^{-1}$, Clayton, Geballe, \& Bianchi 2003),
with the exception of Mira~B (250 and 400 km~s$^{-1}$, Wood, Karovska, \& Hack 2001,
Wood, Karovska, \& Raymond 2002) and V854~Cen (390 km s$^{-1}$, Clayton et al. 1993). 
It is, however, much smaller than those of jets seen in symbiotic objects (e.g.,
$\sim$6000 km~s$^{-1}$ in MWC560, Tomov et al. 1990).
In addition, a change in the line shape in each of the Balmer lines over a period of
75 days is evident.  The blue edge of the absorption features shifted redward  by
200 km~s$^{-1}$, while the red edge did so by only 50 km~s$^{-1}$.

\subsection{Energetics of the Violet Emission}
Our discoveries of the Balmer and Paschen continua, the former being
higher than the latter, the P Cygni-type Balmer series lines, and
their line profile variability all suggest that BM~Gem is associated
with a compact ionized gas region that is accelerated to form a rather
spherical, high velocity outflow.
No mechanism is known for a single cool luminous carbon star to produce an ionized
gas region as well as a high velocity outflow.
They can, however, be accounted for if we introduce an unseen,
low-mass companion that captures matter in the stellar wind from the primary
to give rise to an accretion disk.
This hypothesis is partially supported by similar spectral features observed in
Mira~B ($o$~Cet~B) (see \S 4), which is a low-mass companion to
the AGB star Mira~A ($o$~Cet~A)  and is considered to be associated
with an accretion disk
(Joy 1926, 1954;  Deutsch 1958;  Warner 1972;  Reimers \& Cassatella 1985).
Low-mass companions with an accretion disk have also been suggested for
R Coronae Borealis stars, which show outflows with similar high velocities
and \ion{He}{1} lines that require excitation sources
(Rao et al. 1993; Clayton et al. 2003).

Formation of an accretion disk has proven to be robust around a companion star to
a mass-losing giant in various detached configurations (Mastrodemos \& Morris 1998, 1999).
The inner part of the accretion disk must be heated by the released gravitational
potential energy from the accreting matter.
A hot ionized gas region should form between the innermost region of the
disk and the surface of the companion.
The accretion phenomenon may also be responsible for the discovered high velocity,
variable outflow, although the details of the acceleration mechanism
is not yet settled for this type of outflow.
Alternatively, the outflow may be due to dust formation in the circumcompanion region,
analogous to the scenario proposed for RCB stars by Clayton (1996).
Dust grains would be accelerated through radiation pressure, dragging gas to form
the observed high velocity outflow. The ionization could then be due to collisional
ionization in the flow. In this case, to be compatible with the observed broad
absorption core that extends from the systemic velocity to the terminal velocity
in the P Cygni profile, the ionization must occur before the gas acceleration
is completed. Otherwise the absorber in front of the continuum source has 
terminal outflow velocity and a much narrower absorption component appears
near the terminal velocity. 
In either case of an outflow powered by accretion or driven by radiation
pressure on dust grains, the observed spectral features favor the presence
of a companion to BM~Gem.

The companion should be either a white dwarf or a low-mass dwarf
star because hypotheses invoking a luminous companion have been rejected
by previous studies (Noguchi et al. 1990; Lambert et al. 1990;
Chan \& Kwok 1988; Lloyd-Evans 1990).
The gas expansion velocity provides a hint to distinguish the candidates
for the companion.
White dwarfs have escape velocities on the order of several thousand~km~s$^{-1}$,
while those of low-mass dwarf stars are of the order of several
hundred~km~s$^{-1}$. If gas acceleration occurs near the surface of
the secondary and the flow is aligned with the sight line, the latter is
the case for BM~Gem. However, the former could also be the case if the acceleration
takes place at some point distant from the white dwarf,
as suggested by Warner (1972),
or the flow has a narrow opening angle and is markedly inclined with
respect to the sight line.  Below, we examine whether such a binary and
accretion hypothesis is plausible energetically.

We first attempted to determine the absolute magnitudes of BM~Gem at the observed
wavelengths by finding the distance to BM~Gem. Although it exists,
the Hipparcos parallax of BM~Gem (Table 2) is unreliable.
Claussen et al. (1987) gave 1.51 kpc assuming that carbon
stars have a constant absolute K-band magnitude of -8.1.
We made another estimate by comparing the near-infrared flux
densities of BM~Gem with those of Y~CVn, as they are $^{13}$C-rich (J-type) carbon
stars resembling each other in the spectral characteristics in the optical
and near-infrared (Barnbaum 1991; Ohnaka \& Tsuji 1999, Yamamura et al. 2000),
which may indicate they have similar intrinsic properties.
Y~CVn is the only J-type star that has a relatively reliable
Hipparcos parallax (Table 2) of 218 pc with 16\% uncertainty.
It is evident from the data shown in Table 2 that the differences between Y~CVn and BM~Gem in
$I$ through $L$ bands are quite constant with a simple mean of 3.72,
which indeed shows the two stars to be similar and gives a distance to BM~Gem
of 1.21 kpc.
In addition, the distance of 1.51 kpc reported by Calussen et al. may be reduced
to 1.14 kpc, considering the distance they gave of 0.29 kpc for
Y~CVn should be readjusted to 0.218 kpc.
Hence, we adopted 1.2 kpc with a conservative uncertainty
factor of 1.2 as a reasonable estimate for the distance to BM~Gem.
The total luminosity then becomes
5.4$\times 10^{3}~~L_{\odot}$ (cf. Groenewegen et al. 1992),
which implies that they are on the AGB even with the distance uncertainty.

Interstellar extinction toward BM~Gem ($l=193\fdg2$ and $b=17\fdg2$)
at 3650 {\AA} ($= U$ band), $A_{3650}$, is estimated to be at most
0.6 mag. It is obtained by
\begin{equation}
A_{3650}=\alpha\ N(HI) (N(H)/E(B-V))^{-1} (A_{V}/E(B-V)) (A_{3650}/A_{V}),
\end{equation}
where $N(HI)$ is the column density of atomic neutral hydrogen,
$\alpha$ is a conversion factor from $N(HI)$ to the total column density
of neutral hydrogen $N(H)$ including \ion{H}{1} and H$_{2}$,
$E(B-V) = A_{B} - A_{V}$, and $A_{B}$ and $A_{V}$ are the extinction at
$B$ band and $V$ band, respectively.
We read $N(HI)=7\times10^{20}~{\rm atoms}~{\rm cm}^{-2}$
in the direction of BM~Gem from Heiles (1975).
Other quantities are found in Cox (2000) and references therein:
$N(H)/E(B-V)=5.8\times 10^{21}~{\rm atoms~cm^{-2}~mag^{-1}}$ (Bohlin, Savage, \& Drake 1978);
$R_{V}=A(V)/E(B-V)=3.1$, a standard value for diffuse interstellar matter;
$A_{3650}/A_{V}=1.56$ for $R_{V}=3.1$ (Cardelli, Clayton, \& Mathis 1989).
Then, we find $A_{3650} = 0.6 \alpha$, which should be an upper limit
because the column denisity that Heiles (1975) gave is an upper
limit to BM Gem, which lies somewhere between the boundary
of the HI gas distribution in that direction and us.
Since there are no significant molecular clouds found
in the direction of BM Gem (e.g., Dame, Hartmann, \& Thaddeus 2001),
$\alpha$ should be close to unity and the use of $R_{V}=3.1$ should be justified.
Therefore we regard $A_{3650} = 0.6$ as an upper limit.
  We have also obtained another estimate of $A_{3650} = 0.3$ for the interstellar
extincion toward BM~Gem, using $E(B-V)=0.067$ read from
Schlegel, Finkbeiner, \& Davis (1998) by way of NASA/IPAC Infrared Science Archive,
$R_{V}=3.1$, and $A_{3650}/A_{V}=1.56$.
This is another upper limit because BM Gem must lie somewhere between
the boundary of the dust distribution in the sight line and us.
  Taking these into account we adopt the 0.3 mag directly from
far-inrared dust observations as a nominal value, the 0.6 mag from the \ion{H}{1}
observations as a maximum value, and no extinction as a minimum value,
for $A_{3650}$ to BM~Gem, namely, $A_{3650}=0.3\pm0.3$.
Then we find the absolute magnitudes at
$U$ ($M_{U}$) and 4000 {\AA} ($M_{4000}$) to be
5.2$^{+1.0}_{-1.2}$ and 5.2$^{+1.1}_{-1.3}$, respectively.
These are obtained from the observed magnitudes of $M_{U}$=15.9$^{+0.45}_{-0.65}$
and $M_{4000}$=15.9$^{+0.55}_{-0.75}$, the adopted distance of 1.2 kpc,
its uncertainty factor of 1.2, and the same interstellar extinction
correction of -0.3$\pm$0.3 mag to both of them.
The distance and uncertainty factor corresponds to -10.4$\pm$0.2~mag.

The violet continuum is not simply explained by the photosphere
of a postulated companion. For the case of a dwarf companion,
the probable range of $M_{4000}$ between +6.3 and +3.9
corresponds to a photosphere with a spectral type between
late G and late F, which should show numerous strong absorption
lines in the violet region. The violet continuum of BM~Gem is,
however, featureless except for the \ion{Ti}{1} emission lines due to the primary
carbon star and the Balmer series lines due to the outflow, as mentioned
in section 3.1. The featureless continuum requires that
the companion's photosphere is smeared by veiling with/without obscuration
and contributes only a small fraction of the observed violet continuum.
Considering the S/N ratios in the original high-resolution
spectra of $\sim$15, we should be able to detect absorption
lines of which central depths are as weak as 20\% of
the continuum level at 3 $\sigma$ confidence.
In this violet region, there are many absorption lines
the central depths of which are as strong as 80\% of
the continuum level in late F through late K dwarfs. 
For such strong lines to appear weaker than
the 3 $\sigma$ upper limit of 20\%, the veiling should be
at least 4 times the companion's photosphere.
Possible cases are an F- or G-type dwarf obscured by continuous
absorption and covered with veiling and a dwarf later than
G-type either obscured or not by continuous absorption
and covered with veiling.
Obscured B- or A-type photosphere is not plausible for the violet
continuum because we do not see the Balmer Jump typical for such
spectral types and because the companion should be less
massive than the primary carbon star.
The observed violet continuum also cannot be explained by the photosphere
of a white dwarf companion, because the $U$ band brightness, $M_{U}$,
of a DB white dwarf with effective temperature of 25000~K is 9.1
(Allen 1976), and those classified as DA and later are less luminous.
Therefore, most of the observed violet flux in BM~Gem
should have an origin other than the photosphere of the
assumed companion, which is consistent with our accretion
hypothesis.

The total flux of the observed continuum emission
$F^{cont}=\int_{0}^{\infty}F^{cont}_{\nu}d\nu$ is approximately
obtained by $\nu_{0} F^{cont}_{\nu_{0}}$, where $\nu$ and
$F^{cont}_{\nu}$ mean frequency and the flux of the continuum
at $\nu$ per unit frequency range (flux density)
and $F^{cont}_{\nu_{0}}$ and $\nu_{0}$ are their typical values.
We adopt the frequency at $Ly~ \alpha$ (1216 {\AA}), $\nu_{Ly \alpha}$
for $\nu_{0}$  and the flux density at $\nu_{Ly \alpha}$, $F^{cont}_{\nu_{Ly \alpha}}$
for $F^{cont}_{\nu_{0}}$ to obtain $F^{cont}\sim\nu_{Ly \alpha} F^{cont}_{\nu_{Ly \alpha}}
=\lambda_{Ly \alpha} F^{cont}_{\lambda_{Ly \alpha}}$, where $\lambda_{Ly \alpha}$
is the wavelength of ${Ly~ \alpha}$  and $F^{cont}_{\lambda_{Ly \alpha}}$ is
the flux of the continuum per unit wavelength range at $Ly~ \alpha$ wavelength.
Here, we assume $F^{cont}_{\lambda_{Ly \alpha}}$ is approximated by
$F^{cont}_{\lambda_{BL^{-}}}$,
which is the flux of the continuum per unit wavelength range
immediately shortward of the Balmer limit at 3646 {\AA}, for which we obtained
3.6$\times$10$^{-15}$ erg~cm$^{-2}$~s$^{-1}$ {\AA}$^{-1}$ in \S 3.1.
These approximations should be valid because model spectra of
ionized gas show that the continuum emission diminishes rapidly
toward higher frequencies with respect to $Ly~ \alpha$ line
and because the level of the continuum flux per unit wavelength
from an ionized gas at $Ly~\alpha$ wavelength is similar to that
just shortward in wavelength of the Balmer limit
(e.g., Harrington, Lutz, \& Seaton 1981; Pottasch et al. 1981).
Then, the product $\nu_{Ly \alpha} F^{cont}_{\nu_{Ly \alpha}}$ should
give a reasonable estimate of the total flux. 
The above approximations, however, may overestimate the total flux,
and another uncertainty of (-0, +1) in magnitude should be added.

Now, assuming spherical symmetry of the radiation field,
the gas radiative luminosity, $L_{g}$, can be estimated as
\begin{equation}
L_{g} \simeq 4\pi D^{2} \nu_{Ly \alpha} F^{cont}_{\nu_{Ly \alpha}}
=8\times 10^{32}~{\rm erg~s^{-1}}
\sim0.2~L_{\odot},
\end{equation}
\noindent
where $D$ denotes the distance to BM~Gem.
The total uncertainty range has a magnitude of (-1.2, +2.0) 
by summing those discussed previously, or  0.03--0.6~$L_{\odot}$.
$L_{g}$ in this range is
always larger than the gas kinematic luminosity,
(1/2) $\dot{m}_{out}V_{e}^{2}$, and thermal luminosity, $(3/2) nkT 4\pi R^{2}V_{e}$,
and dominates the total luminosity as far as
$\dot{m}_{out} \leq 10^{-9} M_{\odot}~{\rm yr^{-1}}$ and $T \leq 10^{6}$ K.
Here, $\dot{m}_{out}$, $V_{e}$, and $n$ are the mass ejection rate,
outflow velocity, and total particle number density, respectively,
of the high velocity outflow. $k$, $T$, and $R$ are
the Boltzmann constant, temperature of the gas, and radius under
consideration from the center of the outflow source, respectively.

Maximum energy input, $L_{a}$, expected from the mass-accretion by a companion,
is constrained by the gravitational potential energy release from the
accreting matter, i.e.,
\begin{equation}
L_{a}=G(M_{2}\dot{M}_{2})/r
=0.16~(M_{2}/0.5)(\dot{M}_{2}/10^{-10}) (r/0.01)^{-1}~L_{\odot}, 
\end{equation}
where $G$, $\dot{M}_{2}$, $M_{2}$, and $r$ are the gravitational constant,
the companion's mass accretion rate in $M_{\odot}~{\rm yr^{-1}}$,
its mass in $M_{\odot}$, and its radius in $R_{\odot}$, respectively.
In this equation $M_{2}/r$ is almost constant and near unity
for dwarf stars of type late F to late K,
and is 20--70 for white dwarfs (Allen 1976).
For a white dwarf of 0.5 $M_{\odot}$ with 0.01 $R_{\odot}$ (Allen 1976),
an accretion rate of $(0.2-4)\times 10^{-10}M_{\odot}~{\rm yr^{-1}}$ can afford the observed
continuum luminosity of $(0.03-0.6)~L_{\odot}$,
whereas a K5 dwarf of 0.69 $M_{\odot}$ and 0.74 $R_{\odot}$ (Allen 1976)
can give rise to the observed luminosity when the mass accretion rate
is $(0.1-2)\times 10^{-8}M_{\odot}~{\rm yr}^{-1}$.
It is not self-evident if the latter accretion rate is feasible for
a dwarf companion in a binary system compatible with the observations,
given the current mass loss rate of $3\times 10^{-7}M_{\odot}~{\rm yr}^{-1}$
in BM~Gem (Kahane et al. 1998).

According to Warner (1972), the Bondi-Hoyle type mass accretion rate is written as
\begin{eqnarray}
\nonumber
&&\dot{M}_{2} = 
(1/2) G^{2} \dot{M}_{1} M_{2}^{2} [(v_{rel}^{2}+c^{2})^{3/2}v_{e}]^{-1} d^{-2} \\
&&~~~~=3.4\times 10^{-9}(\dot{M}_{1}/10^{-7})~(M_{2}/0.5)^{2}
[{(v_{rel}^{2}+c^{2})^{3/2}v_{e}} /(7.5)^{4}]^{-1}~(d/30)^{-2} ~~M_{\odot}~{\rm yr}^{-1},
\end{eqnarray}
where $v_{e}$, $v_{rel}$, and c are the mass outflow velocity from the primary
carbon star, relative velocity of the companion to the primary outflow,
and speed of sound of the material, respectively, all in km~s$^{-1}$,
and $\dot{M}_{1}$ and $d$ are the mass loss rate of the primary in $M_{\odot}~{\rm yr^{-1}}$
and binary separation in AU, respectively.
Substituting equation~(4) in equation~(3) we find,
\begin{eqnarray}
&&\nonumber L_{a}=(1/2)G^{3}\dot{M}_{1}M_{2}^{3}  [(v_{rel}^{2}+c^{2})^{3/2}v_{e}rd^{2}]^{-1}\\
&&~~~=5.2~(\dot{M}_{1}/10^{-7})~(M_{2}/0.5)^{3}
[{(v_{rel}^{2}+c^{2})^{3/2}v_{e}} /(7.5)^{4}]^{-1}~(d/30)^{-2}~(r/0.01)^{-1} ~L_{\odot}.
\end{eqnarray}
The speed of sound of the material can be neglected here as it is likely that
the wind is flowing supersonically.
Here, we may write $v_{rel}= ( v_{e}^{2} + v_{orbit}^{2} )^{1/2} $, where $v_{orbit}$
is the orbital velocity of the companion
(Warner 1972; Jura \& Helfand 1984).

If we assume BM~Gem is in a binary system consisting of a carbon star of
1.5~$M_{\odot}$ (Claussen et al. 1987; Groenewegen et al. 1992)
and a dwarf of 0.5~$M_{\odot}$, in a circular orbit with a separation of 30 AU,
which is one likely configuration, then the orbital velocity of the companion
about the primary becomes 7.7~km~s$^{-1}$. The outflow
velocity of the primary carbon star is 7.5~km~s$^{-1}$ and the current
mass loss rate is $\sim3\times 10^{-7}~M_{\odot}~{\rm yr}^{-1}$ in BM~Gem
(Kahane et al. 1998). Then, equation~(5) gives an accretion luminosity of
0.08~$L_{\odot}$,
which is at least compatible with the observed luminosity of BM~Gem
when its uncertainty range is taken into account.
In their smoothed particle hydrodynamic calculations, 
Mastrodemos \& Morris (1999) found that the Bondi-Hoyle accretion is not
as efficient as initially thought. The obtained efficiency ($\dot{M}_{2}$/$\dot{M}_{1}$)
spreads over a range between 0.1\% and 10\% for the cases they
examined that are compatible with the configurations under consideration here.
The efficiencies they found differ not by an order but by a factor
from those obtained using equation~(4) for the same parameter sets.
A dwarf companion is thus at least compatible with the observations.

\section{DISCUSSION}
In the previous section, we showed that our results were compatible with
BM~Gem being accompanied by a low-mass, low-luminosity companion with an
accretion disk, for which both a main-sequence star and a white dwarf are viable.
In this section, we further consider the characteristics of the BM~Gem system.

The observed UV-optical (violet) spectral features of BM~Gem nearly
parallel those of Mira ($o$ Cet). Mira is known to consist of a long
period variable on the asymptotic giant branch, Mira~A 
and a low-mass, low-luminosity companion separated by about $0\farcs6$ ($\sim$70~AU),
Mira~B 
(Karovska et al. 1997 and references therein).
Joy (1926, 1954) reported the first detection of a UV-optical continuum and of complex
profiles of Hydrogen Balmer lines with emission and absorption cores in Mira~B.
Figure~8 of Reimers \& Cassatella (1985) showed that both Balmer
and Paschen continua existed, and the level of the former was higher than that of the latter.
The Balmer lines were found to be in P Cygni-type and their profiles were
shown to be highly variable on a time scale similar to that found for BM~Gem
(Joy 1926, Yamashita \& Maehara 1977), although the lines were not in P Cygni-type
in Reimers \& Cassatella (1985).
The P Cygni profiles indicated that
the outflow velocity was as large as 400~km~s$^{-1}$ (Wood et al. 2002).
Warner (1972) examined the energetics of Mira~B and reported that its total luminosity
of $\sim0.2~L_{\odot}$ could be accommodated by the Bondi-Hoyle type accretion
of Mira~A's wind to Mira~B, if the mass loss rate from Mira~A is greater than
0.8$\times10^{-7}~M_{\odot}$~yr$^{-1}$.  The actual mass loss rate of Mira~A is
$\sim3\times10^{-7}~M_{\odot}$~yr$^{-1}$ (Ryde \& Sch$\ddot{\rm o}$ier 2001),
which supports the Bondi-Hoyle type accretion.
Jura \& Helfand (1984) also found the UV-optical luminosity of Mira~B to be
0.2 (0.05--1)~$L_{\odot}$.
The above parallels between BM~Gem and Mira in the spectral features as well
as the UV-optical luminosity support the presence of
a companion to BM~Gem.

Based on the X-ray luminosity as well as of the outflow velocity of $\sim$400~km~s$^{-1}$,
the companion of BM~Gem may not be a white dwarf but a dwarf.
Here, we again present a parallel discussion with Mira.
A very low X-ray luminosity, which is $\sim$10$^{-3}$ of the UV-optical luminosity,
in the Mira system led Jura \& Helfand (1984) to conclude that Mira~B was not a white
dwarf but a dwarf. They argued that the X-ray luminosity
would be comparable to the UV-optical luminosity if the companion was a white dwarf.
Reimers \& Cassatella (1985) noted that there was no direct evidence for
the presence of a hot white dwarf companion to Mira~A, although they favored
a white dwarf companion. Based on a similar discussion, Ireland et al. (2007)
recently concluded that Mira~B is a K5 dwarf of 0.7~$M_{\odot}$.
If Mira~B is indeed a dwarf, then the X-ray luminosity of BM~Gem would
be as small as that of Mira in the case of a dwarf companion but could be
10$^{3}$ times as much for a white dwarf companion.
Mira has an X-ray photon flux of 7.6$\times10^{-3}$~counts~s$^{-1}$ in
the 0.1--2.4~keV energy band in the second ROSAT source catalog of
pointed observations (ROSAT Consortium 2000).
The X-ray flux from BM~Gem in the same energy band could be
as much as $\sim$0.08~counts~s$^{-1}$ for a white dwarf companion,
with the difference in their distances of a factor 10 taken into account.
It should have been detected in the ROSAT all-sky survey (Voges et al. 1999, 2000),
the detection limit of which was about 0.05~counts~s$^{-1}$,
which is actually not the case.
This argues for a dwarf as the companion of BM~Gem.
However, that both Mira and BM~Gem possess
a white dwarf companion cannot be excluded completely,
because there may exist a mechanism that suppresses
the X-ray luminosity arising from the accretion process
and one to accelerate the outflow from some distant point
from the white dwarf simultaneously.
The latter mechanism is favorable for the picture suggested
for Mira~B by Warner (1972).
It is thus difficult to choose exclusively between
a dwarf and a white dwarf as the companion, although the data favor a dwarf.

The binary separation of the BM~Gem system is only loosely constrained.
The separation is found to be (6--0.3)$\times 10^{1}$~AU for a dwarf companion,
while that for a white dwarf companion is (5--1)$\times 10^{2}$~AU to reproduce
the observed luminosity of 0.03--0.6~$L_{\odot}$ using equation~(5).
We adopted $\dot{M}_{1}=3\times10^{-7}$ and
$v_{e}=7.5$ and assumed that $M_{1}$ (the mass of the
primary in $M_{\odot}$)=1.5, $M_{2}=0.5$, $r$=0.65 (dwarf) or
0.016 (white dwarf),  and the system is in a circular orbit.
Further, for any combination of $M_{1}$ between 1.0 and 2.0 and
$M_{2}$ between 0.3 and 1.2 but $M_{1} > M_{2}$ (dwarf) or
between 0.3 and 0.7 (white dwarf), there are upper limits of 
210~AU and 930~AU for a dwarf and
a white dwarf companion, respectively,
where realistic $r$ corresponding to $M_{2}$ (Allen 1976,  Lang 1999) is used.
As equation~(3) gives an upper limit and equation~(4) is suspected to
overestimate the accretion rate (Mastrodemos \& Morris 1999),
the derived upper limits for the separation are relatively stringent.
Mastrodemos \& Morris (1999) and Soker \& Rappaport (2000) suggested from
theoretical considerations that the separation should be $\lesssim$30~AU to
form an accretion disk around a companion, irrespective of whether it is a dwarf or
a white dwarf, in the detached binary configurations under consideration here.
The observation that Mira~B, which is at least 70~AU from Mira~A,
is associated with an accretion disk despite the moderate mass loss rate
of Mira~A of $\sim3\times10^{-7}~M_{\odot}$~yr$^{-1}$
suggests that the current hydrodynamic simulations do not constrain
the upper-limit of the separation very well.
As BM~Gem has been shown to be a system resembling Mira,  its binary separation
could be as large as 70~AU.
Ohnaka et al. (2006) derived an upper limit of $\sim$60--80~AU
for the diameter of the inner dust-free region in the silicate carbon star IRAS08002-3803
from mid-infrared interferometry, which at the same time gives an upper limit for
the postulated binary separation of $\sim$80~AU. However, they
suggested that silicate carbon stars with an optically thin dust reservoir, to which
BM~Gem belongs, have binary separations wider than those of silicate
carbon stars with an optically thick reservoir, to which IRAS08002-3803 belongs.
In addition, if we pose a constraint that we do not see a Roche-Lobe overflow,
then we find the binary separation should be larger than 1.8 AU for any combination of
mass ratio between 1 and 7 and carbon star radius between 1 and 1.5 AU
using the formulation by Paczy\'nski (1971).

Luminosity of the silicate dust in BM~Gem, $L_{d}$ is estimated as
\begin{equation}
L_{d}=4\pi D^{2} \int F^{silicate}_{\nu}d\nu
\sim 4\pi D^{2} (1/5)\nu_{9.7 \mu m} F^{silicate}_{\nu_{9.7 \mu m}}
= 4\times 10^{35}~{\rm erg~s^{-1}},
\end{equation}
\noindent
where $F^{silicate}_{\nu}$, $\nu_{9.7 \mu m}$,
and $F^{silicate}_{\nu_{9.7 \mu m}}$ are the flux density due to silicate dust,
frequency at 9.7 $\mu$m, and flux density due to silicate dust at 9.7 $\mu$m peak, respectively.
We assume the emission is isotropic, the feature extends over 8-12 $\mu$m,
and has a flux density of about 1.2$\times 10^{-11}$ W m$^{-2}$ at the peak,
which is read from the ``IRAS Low Resolution Spectrograph'' atlas
(Joint IRAS Science Working Group 1986). The factor 1/5 is a roughly determined
correction factor for substituting the integration with the multiplication
obtained by a simple calculation of (1/2)(12 $\mu$m-8 $\mu$m)/(9.7 $\mu$m).
The dust luminosity is then $\sim110~L_{\odot}$.
The accretion process discussed here cannot be the energy source
of the silicate emission lines.
The energy must be supplied by the radiation from the primary carbon star
of $\sim 5 \times 10^{3}~L_{\odot}$ (see \S 3.2).
The dust luminosity indicates that about 2\% of the
total stellar luminosity is captured by the dust
grains responsible for the silicate features. 
Barnbaum et al. (1991) found similar values of 1--2 \%
in BM~Gem and V778~Cyg.

Silicate emission features have been speculated to arise from a circumstellar
disk around a companion as well as from a thickened circumbinary disk
(Lloyd-Evans 1990).
Engels \& Leinert (1994) inferred a minimum radius for a circumbinary
molecular reservoir of 45~sin~$i$~AU, where $i$ denotes the inclination
of the reservoir, which is 90$^{\circ}$ when seen edge-on, for V778~Cyg and
EU~And, assuming a 1~$M_{\odot}$ primary and circular Keplerian motion
of the reservoir. They based their inference on constancy better than
0.06~km~s$^{-1}$~yr$^{-1}$ of the radial velocity of the water maser lines.
Kahane et al. (1998) and Jura \& Kahane (1999) detected
a very narrow emission component in the mm-wave CO lines
in BM~Gem and EU~And. They interpreted the narrow component as due
to a circumbinary reservoir in Keplerian motion formed by some binary
interaction with an unseen companion. They argue that the oxygen-rich
material responsible for the silicate emission is stored in
the long-lived circumbinary reservoir of 100--1000 AU size, which was
formed when the primary star was an oxygen-rich mass-losing star.
This picture is further reinforced by the study of Red Rectangle by Waters et al. (1998)
who noted the similarity to silicate carbon stars of Red Rectangle,
which possesses a circumbinary disk showing crystalline silicate features,
an extended carbon-rich outflow, and narrow CO emission lines.
The circumbinary reservoir of 100--1000 AU size is compatible with
both the binary separation of (0.3--6)$\times 10^{1}$~AU
(as well as a stringent upper limit of 210~AU) derived
for a dwarf companion to BM~Gem and the separation of (1--5)$\times 10^{2}$~AU
(as well as a stringent upper limit of 930~AU)
for a white dwarf companion.

However, Yamamura et al. (2000) found that silicate dust grains responsible for the emission
features in V778~Cyg should have temperature between 600 and 300 K, which
implies their location to be between about 25 and 100 AU (12--50 stellar radii)
from the primary.
They concluded that a circumstellar disk around an invisible companion should
be the reservoir of oxygen-rich material, because they found it difficult to
locate a circumbinary disk in the vicinity of the companion's orbit.
Dust grains there will be swept outward in less than one orbital period
by radiation pressure from the primary.
They speculated that the silicate features originate from oxygen-rich material
continuously blown out by radiation pressure of the primary from the
circumcompanion disk that was built when the primary was an oxygen-rich
mass-losing star.
Furthermore, Ohnaka et al. (2006) suggested that silicate carbon stars may be
classified into two groups: one with an optically thick circumbinary dust reservoir
with a smaller binary separation, and another with an optically thin dust outflow
from the companion's disk with a wider binary separation. Following their criteria,
BM~Gem belongs to the latter group, and thus may show dust outflow from
the companion, possibly extending out in a region at 50--100 AU from the primary. 
Therefore, the region at 30--100~AU is a likely location of the circumstellar 
molecular/dust reservoir according to both Yamamura et al. and Ohnaka et al.
This location is compatible with the binary separation for a dwarf companion,
but not with the separation for a white dwarf companion.

Finally, it should be noted that all of the silicate carbon stars
examined spectroscopically are known to be J-type
stars (Lloyd-Evans 1990, 1991), while the opposite is not true.
It has been argued that J-type carbon stars are the direct descendants of
R-type carbon stars (Lloyd-Evans 1990; Lambert et al. 1990), which form
through a mechanism unrelated to the third dredge-up, as
both groups of stars are $^{13}$C-rich and lack s-process enhancements
in the surface chemical compositions (Utsumi 1985, Dominy 1985).
Abia \& Isern (2000), however, favor another scenario in which
the low-mass J-type stars discussed here form via a combination of
non-standard extra mixing and cool bottom processing (Wasserburg,
Boothroyd, \& Sackmann 1995) early on the AGB,
which is consistent with their luminosities being similar to those
of normal cool carbon stars on the AGB
(Wallerstein \& Knapp 1998; Alksnis et al. 1998).
As there seems to be a connection between the J-type nature
and the silicate features, the presence of a companion to BM~Gem
would suggest a possible connection between binarity and
J-type phenomena, which is worth investigating further.
Radial velocity monitoring of J-type stars would be important,
which could be carried out by observing the \ion{Ti}{1} emission lines
in the violet region as well as numerous photospheric
molecular absorption lines.

\section{CONCLUSIONS}
We observed the violet spectra of the three brightest silicate carbon stars,
BM~Gem, V778~Cyg, and EU~And, using the High Dispersion Spectrograph on the Subaru
telescope, and used a prototypical J-type carbon star Y~CVn for comparison.
Balmer continuum and Paschen continuum emission typical
for ionized gas were found to be prominent in the region shortward of
4000 {\AA} in BM~Gem, while no significant emission was observed
in the same region in V778~Cyg, EU~And, or Y~CVn.
In BM~Gem, Balmer series lines were also detected from H$_{\gamma}$
through H$_{23}$, and they showed P Cygni-type line profiles.
Broad \ion{Ca}{2} K emission with blueward depression was also found,
while the \ion{Ca}{2} H line is missing perhaps because it is absorbed
by hydrogen in the outflow.
The P Cygni profiles give the gas outflow velocity of at least
400 km~s$^{-1}$. Such spectral features have not been observed
in other cool luminous carbon stars.
Furthermore, the P Cygni profiles changed significantly within a period of 75 days,
suggesting a compact geometry of the outflow.
The overall spectral features mimic those of Mira B.
All these features in the cool carbon star BM~Gem
suggest the presence of a companion that gives rise to
an accretion disk, which is responsible for the ionized gas region and
the observed high velocity outflow.

We investigated the energetics of the observed emission assuming
a binary system and Bondi-Hoyle type mass accretion process.
The luminosity of the observed continuum emission is estimated to be
$\sim$0.2 (0.03--0.6)~$L_{\odot}$, while the silicate dust features convey
about 110 $L_{\odot}$, which shows that the silicate dust grains are
heated not by the phenomenon discovered here but by radiation
from the carbon star. We found that the violet continuum luminosity
is accommodated if an accretion rate on the order of
$10^{-8}~M_{\odot}$~yr$^{-1}$  is achieved to a dwarf
companion, while $10^{-10}~M_{\odot}$~yr$^{-1}$ is sufficient to
a white dwarf companion. Although the required accretion efficiency
seems rather high, a main-sequence companion is favored based on
the observed outflow velocity of 400 km~s$^{-1}$ and the
non-detection of X-ray flux in the ROSAT all sky survey.
The possibility of a white dwarf companion, however, cannot be ruled out,
because the outflow could be narrow and inclined with respect to
the sight line or be accelerated at some distant point from the
surface of the white dwarf. 

Based on the above findings combined with those of previous studies, 
we conclude that BM~Gem is associated with a low-mass,
low-luminosity companion giving rise to an accretion disk.
We note that the spectral features and UV-optical luminosity
obtained for the BM~Gem system closely resemble those
of the Mira system, which should represent the actual similarity
between the binary configurations of the two systems.
We derived the binary separations based on the observed UV-optical
luminosity as (0.3--6)$\times 10^{1}$~AU and (1--5)$\times 10^{2}$~AU
for a dwarf and a white dwarf companion, respectively,
assuming the primary mass of 1.5~$M_{\odot}$,
companion mass of 0.5~$M_{\odot}$,
circular orbits, and Bondi-Hoyle type accretion.
We also derived another stringent upper limit of the separation
of 210~AU and 930~AU
for a dwarf and a white dwarf companion, respectively.
The separations for a dwarf are compatible with both the circumcompanion dust
reservoir proposed by Yamamura et al. (2000) and Ohnaka et al. (2006)
and the circumbinary dust reservoir proposed by Kahane et al. (1998) and Jura \& Kahane (1999),
while those for a white dwarf are plausible only for the circumbinary reservoir.
However, it is still difficult to choose between the two scenarios
for the dust reservoir responsible for the silicate emission features in BM~Gem.
A very high angular resolution imaging in the ultraviolet would be interesting
to depict the reservoir through the violet continuum emission scattered
by the silicate grains.
If the silicates are located in a circumbinary reservoir, we will see
a ring-like structure, while we may see a spiral structure if they are
blown out continuously from a circumcompanion disk.
It is also important to carry out high sensitivity, medium
resolution spectroscopy in the ultraviolet of
silicate carbon stars other than BM~Gem to determine whether 
the violet spectral features are common among silicate
carbon stars. Theoretical studies of the formation mechanism 
of the dust reservoir are also required.

\acknowledgments
The authors thank Dr. Geoff Clayton, the referee of this paper, for valuable
comments and careful reading that helped improve the manuscript considerably.
The authors are also grateful to all the staff members of the Subaru telescope.
H.I. and K.N. were supported by a Grant-in-Aid for Scientific Research
(C) (No.13640247) from the Japan Society for the Promotion of Science (JSPS).
H.I. and M.Y. were also partly supported by a JSPS grant for Scientific Research (A)
(No.14204018). This research made use of SIMBAD and VizieR databases, 
maintained by CDS (Strasbourg), France, and of the NASA/IPAC Infrared
Science Archive, which is operated by the Jet Propulsion Laboratory,
California Institute of Technology, under contract with the National
Aeronautics and Space Administration.

\clearpage

\begin{deluxetable}{lllrllr}
\tablecolumns{7}
\tablewidth{0pc}
\tablecaption{Observation Summary}
\tablehead{
\colhead{} & \colhead{RA(h m s)} & \colhead{Dec($^\circ$ ' ")} & \colhead{Exp} & \colhead{Res.} & \colhead{Date} & \colhead{S/N} \\
\colhead{Name} & \colhead{(J2000)} & \colhead{(J2000)} & \colhead{(sec)} & \colhead{($\lambda/\Delta\lambda$)} & \colhead{(UT)} & \colhead{(@4000 {\AA})}}
\startdata
BM~Gem   & 07 20 58.9 & +25 00 07.28 & 3600  & 50,000 & 2001 Jan 29 & $\sim$15~~ \\
    &            &              & 1800  & 50,000 & 2001 Apr 14 & $\sim$15~~ \\
V778~Cyg & 20 36 07.4 & +60 05 26.2  & 2400  & 38,000 & 2001 Jul 30 & $<$1~~ \\
EU~And   & 23 19 58.2 & +47 14 28    & 2400  & 38,000 & 2001 Jul 30 & $<$1~~ \\
Y~CVn    & 12 45 07.8 & +45 26 24.9  &  900  & 50,000 & 2001 Feb 1 & $\sim$1~~ \\
Feige~34 & 10 39 36.7 & +43 06 09.3  &  300  &  9,000 & 2001 Jan 28 & $\sim$80~~ \\
\enddata
\end{deluxetable}

\begin{deluxetable}{lllllrrrrr}
\tablecolumns{10}
\tablewidth{0pc}
\tablecaption{Astrometric and Photometric Data}
\tablehead{
\colhead{} & \colhead{$\pi$}& \colhead{$\sigma_{\pi}$} & \colhead{} & \colhead{} & \colhead{} &
\colhead{} & \colhead{} & \colhead{} & \colhead{} \\
\colhead{Name} & \colhead{(mas)} & \colhead{(mas)} & \colhead{B} & \colhead{V} &\colhead{I} &
\colhead{J} & \colhead{H} & \colhead{K} & \colhead{L}}
\startdata
BM~Gem   & 1.83  & 1.24 & 10.72 & 8.40  & 4.96 & 4.28 & 3.32 & 2.75 & 2.47 \\
V778~Cyg & \nodata  &  \nodata & 13.75 & 10.26 & \nodata   & 5.25 & 4.20 & 3.54 & 3.18 \\
EU~And   & \nodata  &  \nodata & 13.70 & 10.47 & \nodata   & 5.41 & 4.37 & 3.79 & 3.33 \\
Y~CVn    & 4.59  & 0.73 & 8.41  & 5.42  & 1.25 & 0.59 & -0.37 & -0.89 & -1.42 \\
\enddata
\end{deluxetable}


\clearpage
\begin{figure}
\includegraphics[width=12cm,angle=-90,clip]{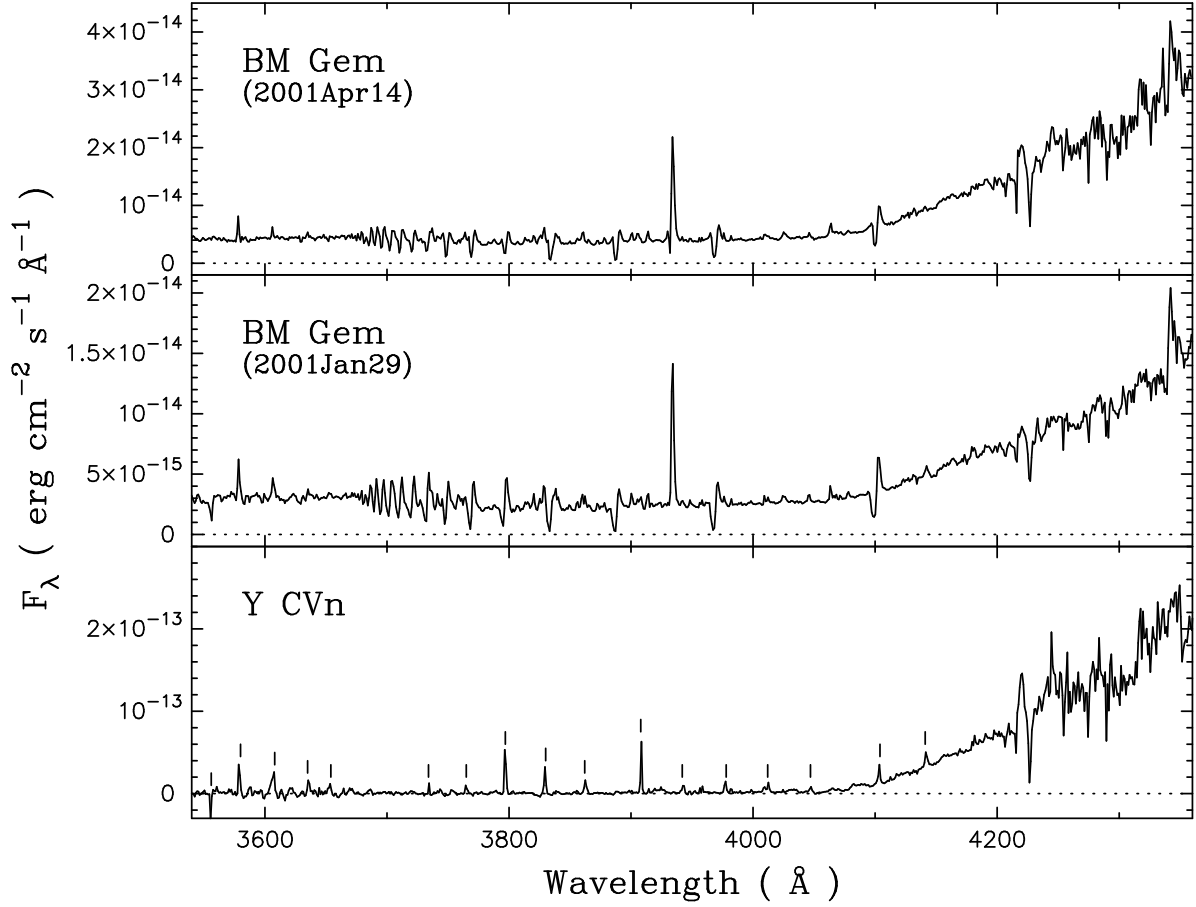}
\caption{Violet spectra of a silicate carbon star, BM~Gem, on
January 29 and April 14, 2001, and a prototypical J-type
carbon star, Y~CVn. All the spectra are binned to 1 {\AA} step.
The dotted horizontal line indicates the zero level.
Note the significant continuum emission,
Balmer lines, and \ion{Ca}{2} K emission in BM~Gem.
The small difference in the continuum level is probably due to
the measurement uncertainty (see text for details). 
Small vertical ticks in the panel of Y~CVn give the
positions of artifacts due to strong red light contamination,
which is also present in BM~Gem but is not prominent because
of the strong continuum (see text for details)
\label{fig1}}
\end{figure}

\clearpage 

\begin{figure}
\includegraphics[width=12cm,angle=-90,clip]{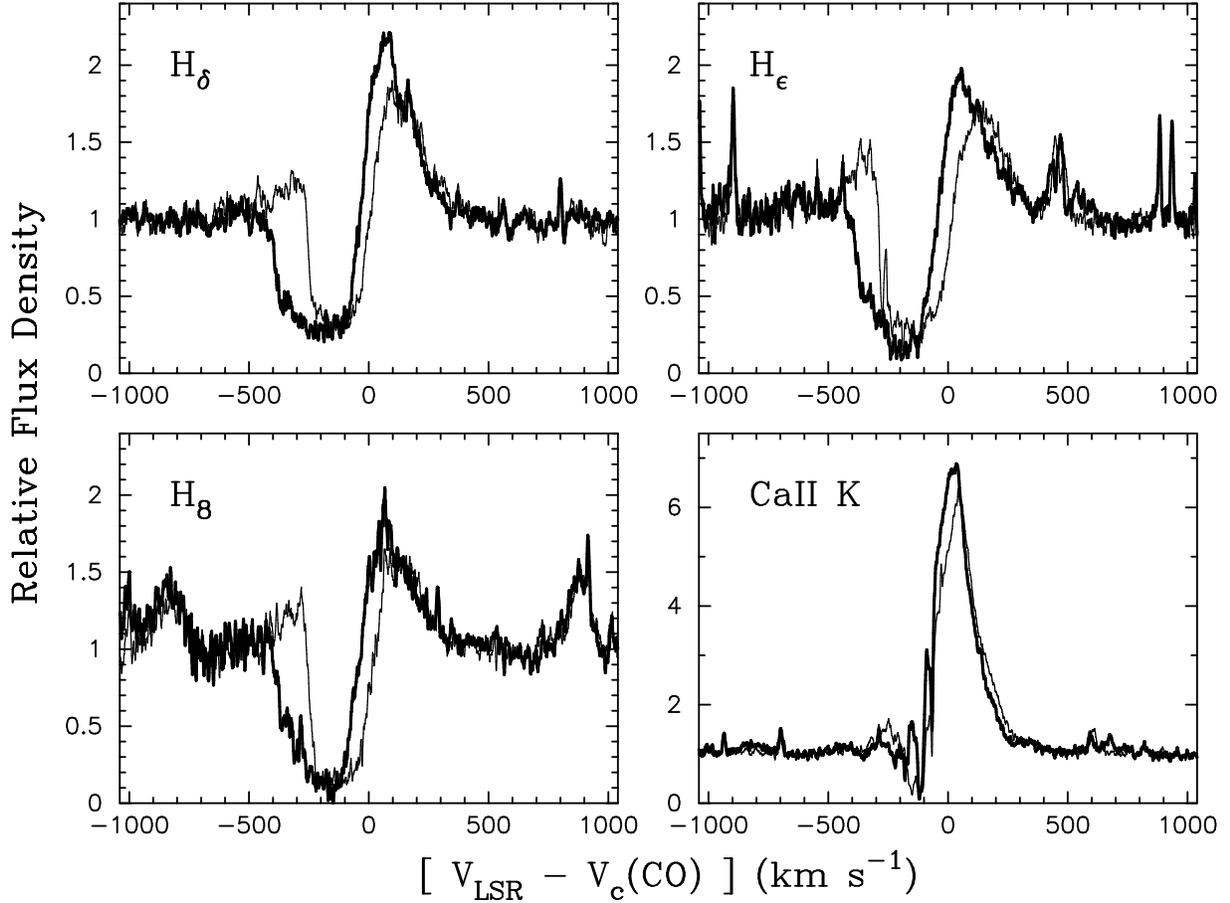}
\caption{
H$_{\delta}$ (top-left), H$_{\epsilon}$ (top-right),
H$_{8}$ (bottom-left), and \ion{Ca}{2} K (bottom-right) lines
in BM~Gem observed on January 29 (thick line) and
April 14, 2001 (thin line), shown with the original resolution
($\lambda/\Delta \lambda$) of 50,000. The spectra are normalized
to the local Paschen continuum. For H$_{\delta}$ the contribution
of the carbon star spectrum was approximately subtracted.
The profiles indicate that the gas expansion velocity in January
was as large as 400 km~s$^{-1}$
\label{fig2}}
\end{figure}

\end{document}